\documentclass[letterpaper,11pt]{article}

\usepackage{dynkin-diagrams}
\usepackage{cite}
\usepackage{amsmath}
\usepackage{amssymb}
\usepackage{mathrsfs}
\usepackage{xypic}

\usepackage{hyperref}
\usepackage{color}
\definecolor{darkred}{rgb}{0.65,0.15,0}
\hypersetup{pdfborder={0 0 0},colorlinks=true,urlcolor=darkred,citecolor=blue,linkcolor=darkred,linktocpage=true}

\usepackage{float} 

\usepackage{geometry}

\newgeometry{vmargin={35mm}, hmargin={35mm}} 

\usepackage{keyval}
\makeatletter
\define@key{setpar}{left}[0pt]{\leftmargin=#1}
\define@key{setpar}{right}[0pt]{\rightmargin=#1}
\define@key{setpar}{both}{\leftmargin=#1\relax\rightmargin=#1}
\makeatother

\newenvironment{narrow}[1][]
  {\list{}{\setkeys{setpar}{left,right}%
     \setkeys{setpar}{#1}%
     \listparindent=\parindent
     \topsep=0pt
     \partopsep=0pt
     \parsep=\parskip}\item\relax\hspace*{\listparindent}\ignorespaces}
  {\endlist}
  
\def\eg{{\it e.g.}}

\def\adj{\hbox{\bf adj}}

\def\EWeight#1#2#3#4{\bigl({}^{\mathstrut}_{#1\mathstrut}{}_{#2\mathstrut}^{#4\mathstrut}{}_{#3\mathstrut}^{\mathstrut}\bigr)}

\def\DWeight#1#2#3{\bigl(\raise2.5pt\hbox{${}_{#1}$}{}^{#2}_{#3}\bigr)}

\def\AAWeight#1#2{\bigl(\raise0pt\hbox{${}^{#1}_{#2}$}\bigr)}

\numberwithin{equation}{section}

\begin{document}

\frenchspacing

%\null\vspace{-28mm}
%
%\includegraphics[height=2cm]{AvancezCHALMERSx2.eps}
%\hspace{2mm}
%\includegraphics[height=1.85cm]{logotype-en.png}
%
%\vspace{-12mm}
%{\flushright Gothenburg preprint \\ 
  %\today\\
%}
%March, 2021\\} %July, 2019\\}
%
%\vspace{4mm}

%\hrule

%\vspace{16mm}

%{\flushright {\tt \today}\\
%{\tt linfty4.tex}\\
%}
%\null\vspace{1cm}

\thispagestyle{empty}

\vfill

\begin{center}
  {\Large \sc \bf Extended geometry of magical supergravities}
%  \\[3mm]
 % {\Large \bf \sc to extended geometry}
 
 \vfill
    
{\large
Guillaume Bossard${}^1$,
Martin Cederwall${}^{2,3}$,
Axel Kleinschmidt${}^{4,5}$,\\[2mm]
Jakob Palmkvist${}^{6}$,
Ergin Sezgin${}^{7}$
and Linus Sundberg${}^8$
}

\vfill

       {\footnotesize ${}^1${\it Centre de Physique Th\'eorique, CNRS, Institut Polytechnique de Paris,
91128 Palaiseau cedex, France}}

\vspace{2mm}
       {\footnotesize ${}^2${\it Department of Physics,
         Chalmers Univ. of Technology, SE-412 96 Gothenburg, Sweden}}

\vspace{2mm}
       {\footnotesize ${}^3${\it NORDITA, Hannes Alfvéns väg 12, SE-106 91 Stockholm, Sweden}}

\vspace{2mm}
       {\footnotesize ${}^4${\it Max-Planck-Institut f\"ur Gravitationsphysik (Albert-Einstein-Institut),\\ 
       Am Mühlenberg 1, DE-14476 Potsdam, Germany}}

\vspace{2mm}
       {\footnotesize ${}^5${\it International Solvay Institutes, ULB-Campus Plaine CP231,
       BE-1050 Brussels, Belgium}}

\vspace{2mm}
       {\footnotesize ${}^6${\it Department of Mathematics,
         \"Orebro Univ.,
 SE-701 82 \"Orebro, Sweden}}

\vspace{2mm}
       {\footnotesize ${}^7${\it Mitchell Institute for Fundamental Physics and Astronomy,\\
       Texas A\&M Univ., College Station, TX 77843, USA}}

\vspace{2mm}
       {\footnotesize ${}^8${\it Department of Physics, Gothenburg Univ., SE-412 96 Gothenburg, Sweden}}

\end{center}

\vfill

\begin{quote}
  
\textbf{Abstract:} 
We provide, through the framework of extended geometry, a geometrisation of
the duality symmetries appearing in magical supergravities.
A new ingredient is the general formulation of extended geometry with structure group of non-split real form.
A simple diagrammatic rule for solving the section constraint by inspection of the Satake diagram is derived.
\end{quote} 

\vfill

%\hrule

%\noindent{\tiny email:
%  martin.cederwall@chalmers.se, jakob.palmkvist@oru.se}

\newpage

\pgfkeys{/Dynkin diagram, edge length=1cm,
fold radius=.6cm, 
root-radius=.11cm,
indefinite edge/.style={
draw=black, fill=white, dotted,
thin}}

\def\fg{{\mathfrak g}}
\def\fh{{\mathfrak h}}
\def\fk{{\mathfrak k}}

\def\gl{\mathfrak{gl}}
\def\sl{\mathfrak{sl}}
\def\so{\mathfrak{so}}
\def\su{\mathfrak{su}}
\def\sp{\mathfrak{sp}}
\def\fe{\mathfrak{e}}
\def\ff{\mathfrak{f}}

\def\RR{{\mathbb R}}
\def\CC{{\mathbb C}}

\def\id{{\mathbb I}}

\def\LL{{\mathscr L}}

\def\tr{\,\hbox{tr}}
\def\diag{\,\hbox{diag}}

\def\ie{\hbox{\it i.e.}}

\def\be{\begin{equation}}
\def\ee{\end{equation}}

\def\bea{\begin{eqnarray}}
\def\eea{\end{eqnarray}}

\def\*{\partial}
\def\Spin{\hbox{Spin}}

\def\KK{{\mathbb K}}
\def\RR{{\mathbb R}}
\def\CC{{\mathbb C}}
\def\HH{{\mathbb H}}
\def\OO{{\mathbb O}}

\def\nn{\nonumber}

\tableofcontents

\section{Introduction and summary}

%We will reformulate the bosonic sectors of the $D=6$ magical supergravities %%%
%\cite{Gunaydin:1983rk,Gunaydin:1983bi,Gunaydin:2010fi} using extended geometry, 
%thus giving a geometric origin to the duality symmetries arising on dimensional reduction.

There exists a special class of supergravity theories in $d=3,4,5,6$, known as magical supergravities \cite{Gunaydin:1983rk,Gunaydin:1983bi,Gunaydin:2010fi} whose symmetries are associated with the remarkable geometries of the magic square of Freudenthal, Rozenfeld and Tits \cite{Tits:1966,Sudbery:1984}. The scalar manifolds arising in all magical supergravities are displayed in Table \ref{tab:Dimension_groups}. The magical theories in $d=6$ are parent theories from which all magical supergravities in $d=3,4,5$ can be obtained by dimensional reduction. The geometries arising in $d=3,4,5$ \cite{Gunaydin:1983rk} were later referred to as very special quaternionic K\"ahler, very special K\"ahler and very special real, respectively. See ref. \cite{VanProeyen:2001wr} for a review.

The use of extended geometry as a means to provide a geometric origin of duality symmetries in string theory and M-theory is well established, see \eg\ refs. \cite{Hull:2007zu,Pacheco:2008ps,Hillmann:2009pp,Berman:2010is,Berman:2011pe,Coimbra:2011ky,Coimbra:2012af,Berman:2012vc,Park:2013gaj,Cederwall:2013naa,Cederwall:2013oaa,Aldazabal:2013mya,Hohm:2013pua,Blair:2013gqa,Abzalov:2015ega,Hohm:2013vpa,Hohm:2013uia,Hohm:2014fxa,Cederwall:2015ica,Bossard:2017aae,Bossard:2018utw,Bossard:2021jix,Bossard:2021ebg,Bossard:2019ksx} for exceptional geometry and 
\cite{Palmkvist:2015dea,Cederwall:2017fjm,Cederwall:2018aab,Cederwall:2019bai,Cederwall:2021xqi}
for the general framework.
The duality symmetries, traditionally arising as an enhancement after dimensional reduction, then become present in the unreduced models, not as global symmetries, but as structure groups of generalised diffeomorphisms.

The present letter aims to fill a gap in the formalism, namely to deal with and interpret duality groups/structure groups of non-split real form. Our main application will be the bosonic sector of the (ungauged) magical supergravities, but the method is generic and can be applied to other models. 
We will thus provide a ``geometrisation'' of the duality symmetries appearing in the magic square. The groups appear as structure groups of extended geometries for different splits of the $6$ dimensions into $n$ ``internal'' and $d$ ``external'' directions, without dimensional reduction. 

A brief recapitulation of magical supergravities is given in Section 2. In Section 3, we recall some basic properties of real forms and Satake diagrams, and also discuss real forms of tensor hierarchy algebras. The latter are used to identify the bosonic fields. 
Section 4 is devoted to the actual construction of the extended geometry, which mimics the formulation of exceptional geometry for $D=11$ supergravity, and to the solution of the section constraint.

\section{Magical supergravities}

Magical supergravities \cite{Gunaydin:1983rk,Gunaydin:1983bi,Gunaydin:2010fi}
are $N=(1,0)$ supergravities in $6$ dimensions, coupled to $n_V$ gauge multiplets and $n_T$ self-dual tensor multiplets, which for particular values of $n_V$ and $n_T$ exhibit enlarged duality symmetry.
This symmetry is $Spin(1,\nu+1)$, $\nu=1,2,4,8$, when the number of multiplets are chosen to be
$n_V=2\nu$, $n_T=\nu+1$. Note that $2\nu$ is the real dimension of a spinor (chiral when $\nu>1$) of $Spin(1,\nu+1)$.
The $\nu+1$ scalars in the tensor multiplet parametrise the coset $SO(1,\nu+1)/SO(\nu+1)$.
In addition to $Spin(1,\nu+1)$ there is also a $U(1)$ for $\nu=2$ and an $SU(2)$ for $\nu=4$, acting on the spinors of
$Spin(1,\nu+1)\simeq SL(2,\KK_\nu)$.
The anti-self-dual tensor $H^0$ in the supergravity multiplet and the $\nu+1$ self-dual ones $H^i$ in the tensor multiplets then combine into $H^I=H^0V_0{}^I+H^iV_i{}^I$, where $(V_0{}^I,V_i{}^I)$ parametrises the scalar coset.
The identity needed for the extra symmetry is the Fierz identity
$\gamma_{a(\alpha\beta}\gamma^a{}_{\gamma\delta)}=0$, valid for spinors of $SL(2,\KK_\nu)$,
$\KK_\nu=\RR,\CC,\HH,\OO$ for $\nu=1,2,4,8$. This can be seen as an identity for elements in the Jordan algebra
$J_2(\KK_\nu)$ of $2\times2$ hermitean matrices. 

%An arbitrary number $n_H$ of hypermultiplets can be added. 

When a magical supergravity is dimensionally reduced to $d<6$ dimensions, the symmetry is further enhanced, leading to the groups in Table \ref{tab:Dimension_groups}, forming a magic square of Lie groups
\cite{Tits:1966,Sudbery:1984}. 
The table may in principle be continued with infinite-dimensional algebras to the right, with a $d=2$ column containing affine extensions of the algebras in $d=3$ column, over-extended Kac--Moody algebras in a $d=1$ column etc. 
The $d=5$ groups are the the structure groups of the Jordan algebras $J_3(\KK_\nu)$ of hermitean $3\times3$ matrices, and the $d=4$ groups the conformal groups of the same algebras.

\begin{table}[H]
    \centering
    \begin{tabular}{c|cccc}
        \rule{0pt}{18pt}$\mathbb{K_\nu}$ & \quad$d=6$ &\quad $d=5$ & $d=4$ & $d=3$\\[6pt] \hline
        \rule{0pt}{18pt}$\mathbb{R}$ & $\frac{SO(1,2)}{SO(2)}$ &\quad\quad $\frac{SL(3,\mathbb{R})}{SO(3)}$ & $\frac{Sp(6,\mathbb{R})}{U(3)}$ & $\frac{F_{4(4)}}{USp(6)\times USp(2)}$\\[6pt]
        \rule{0pt}{18pt}$\mathbb{C}$ & $\frac{SO(1,3)}{SO(3)}$ &\quad\quad $\frac{SL(3,\mathbb{C})}{SU(3)}$ & $\frac{SU(3,3)}{SU(3)\times SU(3) \times U(1)}$ & $\frac{E_{6(2)}}{SU(6)\times SU(2)}$\\[6pt]
       \rule{0pt}{18pt} $\mathbb{H}$ & $\frac{SO(1,5)}{SO(5)}$ &\quad\quad $\frac{SU^*(6)}{USp(6)}$ & $\frac{SO^*(12)}{U(6)}$ & $\frac{E_{7(-5)}}{SO(12)\times SU(2)}$\\[6pt]
       \rule{0pt}{18pt} $\mathbb{O}$ & $\frac{SO(1,9)}{SO(9)}$ &\quad\quad $\frac{E_{6(-26)}}{F_4}$ & $\frac{E_{7(-25)}}{E_6\times SO(2)}$ & $\frac{E_{8(-24)}}{E_7\times SU(2)}$\\
    \end{tabular}
        \caption{\it The cosets, \ie, the structure groups and maximal compact subgroups, for the magical supergravities.}
    \label{tab:Dimension_groups}
\end{table}

Note that the group $Spin(1,9)$ occurring in the octonionic magical supergravity is another real form of $Spin(10)$
than $Spin(5,5)$, the U-duality group for $D=11$ supergravity reduced to $d=6$, and that
the modules of the $1$-form and $2$-form potentials also are ``the same'' in the two cases.
In Section \ref{THAsection}, we will see how these real algebras and modules appear 
in level decompositions of real forms of the same tensor hierarchy algebra over $\CC$.

Already the magical supergravities in $d=6$ will be formulated as extended geometry, where the scalar coset is parametrised as a generalised vielbein on an internal space, however with a section constraint whose solution is a point---the structure group then becomes R-symmetry.

%Do the comparison with reduction of $D=11$ supergravity here.

\section{Real forms and Satake diagrams}

\subsection{Satake diagrams}

We do not aim to give a complete account of Satake diagrams 
\cite{Satake:1960,Araki:1962,OnishchikVinberg} 
and real forms of semi-simple Lie algebras. Rather, some essential features that turn out to be relevant to the present work are described. There are essentially two alternative ways to characterise real forms diagrammatically, Satake diagrams and Vogan diagrams \cite{Knapp:2002}. Roughly speaking, while the Satake diagram describes the deviation from the split real form, the Vogan diagram relates the real form to the compact one.
The Satake diagrams have the advantage that they are in 1-1 correspondence with the real forms.
See also the presentations in refs. \cite{Henneaux:2007ej,Henneaux:2011mm} and in ref. \cite{Keurentjes:2002rc}, which contains examples relevant to the present paper.
Exceptional extended geometry has so far exclusively used structure algebras of split real form. It will become clear, in particular in Section
\ref{SectionSection}, where we solve the section constraint diagrammatically, that the classification using Satake diagrams is much better suited to our purposes.

Let the complex semi-simple Lie algebra $\fg_\CC$ have a Dynkin diagram $\Delta(\fg_\CC)$. A real form $\fg$ of $\fg_\CC$ is a subalgebra over $\RR$, whose complexification is $\fg_\CC$. The complex conjugation of an element $za$, where $z\in\CC$ and $a\in\fg$ is (of course) defined by complex conjugation of $z$, $za\mapsto\bar za$. This defines an (anti-linear) involution 
$\sigma$ on $\fg_\CC$. Conversely, the fixed points of this involution define the real form $\fg\subset\fg_\CC$.
The Satake diagram $\Delta(\fg)$ for the real form $\fg$ encodes the involution $\sigma$, and is a decorated version of
$\Delta(\fg_\CC)$. 

As a preparation, consider $A_1=\frak{sl}(2)$. This complex Lie algebra has two real forms, the compact $\frak{su}(2)$ and the split (maximally non-compact) $\frak{sl}(2,\RR)$. The involution $\sigma$ defining the split real form is the identity involution, and the one defining the compact real form is the Chevalley involution $\sigma$: $e\mapsto-f$, $f\mapsto-e$, $h\mapsto-h$. 
In the split case, the node remains undecorated (white), and in the compact case, the node is colored black.
Any simple Lie algebra has a split real form, defined by the identity involution, whose Satake diagram is identical in appearance to the Dynkin diagram, and a compact real form, defined by the Chevalley involution, whose Satake diagram consists of only black nodes.

There is yet another type of decoration appearing in Satake diagrams, namely arrows. To understand their meaning, consider
the Lie algebra $\frak{sl}(2,\CC)$ as a 
{\it real} Lie algebra. Write an element as $a+ib$, where $a,b\in\frak{sl}(2,\RR)$. In the complexification $\frak{sl}(2,\CC)\otimes\CC$,
where we use another imaginary element $i'$ for the factor $\CC$, we can choose elements
of the forms $a_\pm=P_\pm a={1\over2}(1\pm i\otimes i')a$, projecting on the two parts of $\frak{sl}(2)\oplus\frak{sl}(2)$. The involution corresponding to the real form $\frak{sl}(2,\CC)$ maps $i'\mapsto-i'$, so it interchanges the same basis elements in the two 
$\frak{sl}(2)$'s. Such an involution is denoted by an arrow between the nodes of the two algebras, resulting in the Satake diagram of Figure \ref{SL2CSatake}. Arrows may also appear in a connected diagram.

\vskip4\parskip
\begin{figure}[H]

\begin{center}

\dynkin[involutions=12, edge/.style=white]A{oo}
%\dynkin[involutions={[in=180,out=0,relative]12}, edge/.style=white]A{oo}

\begin{narrow}[both=1cm]
\caption{\it Satake diagram for $\frak{sl}(2,\CC)$.
\label{SL2CSatake}}  
\end{narrow}

\end{center}

\end{figure}

The general rules are as follows:
For a black (compact) node, the involution acts as the Chevalley involution of the corresponding $\frak{sl}(2)$ subalgebra.
For two nodes $i,i'$ connected by an arrow, and unconnected to black nodes, $(e_i,f_i,h_i)\leftrightarrow(e_{i'},f_{i'},h_{i'})$.
For a white (non-compact) node which is not connected to a black node, nor have a connected arrow, the involution acts as the identity on the corresponding 
$\frak{sl}(2)$ subalgebra.
The only complication, and the only action of the involution that can not be immediately read off from the Satake diagram, is the behaviour of the generators associated to a white node, say number $i$, connected to black nodes (which in turn can be connected to further black nodes). The action of the involution $\sigma$ on the $\frak{sl}(2)$ generators is then more complicated. In terms of the induced action of
$\sigma$ on the roots,
 %in spite of belonging to a white node, 
 a simple root $\alpha_i$ corresponding  to an undecorated white node maps to
$\alpha_i+\sum_jc_j\alpha_j$, where the range of the index $j$ is over the group of compact nodes connected (not necessarily directly, but via black nodes) to node $i$. The numbers $c_j$ are positive integers. They must 
be chosen so that the Cartan matrix is invariant (which is obviously impossible if they are zero), and of course so that 
$\sigma^2=1$. If two white nodes (number $i$ and $i'$) are connected with arrows, and in addition both connected via a number of  
black nodes, labelled by an index $j$, one analogously has
$\alpha_i\mapsto\alpha_{i'}+\sum_jc_j\alpha_j$, $\alpha_{i'}\mapsto\alpha_i+\sum_jc'_j\alpha_j$

\begin{figure}[H]
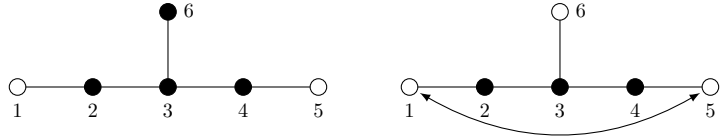


\begin{center}

\dynkin[label,ordering=Dynkin] E{o***o*} \qquad \dynkin[label,ordering=Dynkin,involutions=15] E{o***oo}

\begin{narrow}[both=1cm]
\caption{\it Satake diagrams for $\fe_{6(-26)}$ and $\fe_{6(-14)}$.
\label{E6Satake}}  
\end{narrow}

\end{center}

\end{figure}

We illustrate with two example, of which one appears as one of the structure algebras in magical supergravity,
namely $\fe_{6(-26)}$ and $\fe_{6(-14)}$. The Satake diagrams and the convention for numbering of nodes are given in Figure \ref{E6Satake}.
The Cartan matrix $A$ is
\begin{align}
A=\left(\begin{matrix}
2&-1&0&0&0&0\cr
-1&2&-1&0&0&0\cr
0&-1&2&-1&0&-1\cr
0&0&-1&2&-1&0\cr
0&0&0&-1&2&0\cr
0&0&-1&0&0&2
\end{matrix}\right)
\end{align}
and the two involutions act on the simple roots as
\begin{align}
\sigma_{\fe_{6(-26)}}=\left(\begin{matrix}
1&2&2&1&0&1\cr
0&-1&0&0&0&0\cr
0&0&-1&0&0&0\cr
0&0&0&-1&0&0\cr
0&1&2&2&1&1\cr
0&0&0&0&0&-1
\end{matrix}\right)\;,
\quad
\sigma_{\fe_{6(-14)}}=\left(\begin{matrix}
0&1&1&1&1&0\cr
0&-1&0&0&0&0\cr
0&0&-1&0&0&0\cr
0&0&0&-1&0&0\cr
1&1&1&1&0&0\cr
0&1&2&1&0&1
\end{matrix}\right)
\end{align}
The diagonal elements of the $\sigma$'s are given by the rules ($+1$ for white, $-1$ for black, $0$ when connected by an arrow). 
Nodes $i,j$ connected by an arrow have $\sigma_{ij}=1$. The remaining non-zero numbers (only present for white nodes connected to black ones, in the first example nodes $1$ and $5$, in the second nodes $1$, $5$ and $6$) are not immediately visible in the diagrams, but they are completely determined 
by the conditions $\sigma^2=1$ and $\sigma A \sigma^t=A$.

\begin{figure}[H]
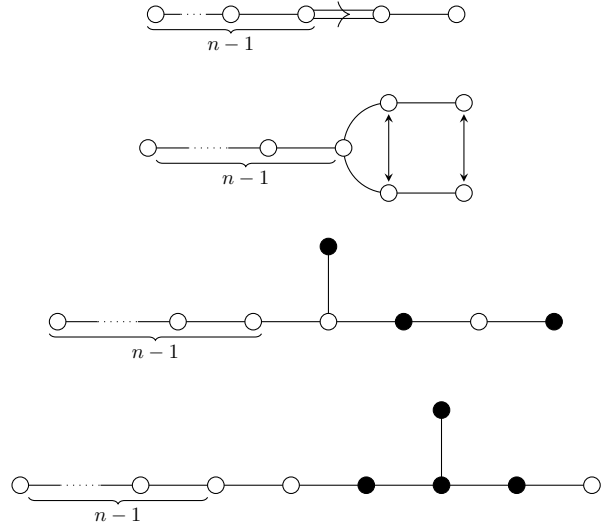


\begin{center}

\begin{dynkinDiagram}[make indefinite edge={0-1}]F[1]{oooo}
\dynkinBrace[n-1]02
\end{dynkinDiagram}

\vspace{12pt}

\begin{dynkinDiagram}[backwards=true,ply=2,make indefinite edge={0-2}]E[1]{oooooo}
\dynkinBrace[n-1]04
\end{dynkinDiagram}

\vspace{12pt}

\begin{dynkinDiagram}[make indefinite edge={0-1}]E[1]{o*oo*o*}
\dynkinBrace[n-1]03
\end{dynkinDiagram}

\vspace{12pt}

\begin{dynkinDiagram}[backwards=true,make indefinite edge={0-8}]E[1]{o****ooo}
\dynkinBrace[n-1]07
\end{dynkinDiagram}

\begin{narrow}[both=1cm]
\caption{\it Satake diagrams 
of the duality groups of the magical supergravities with $n=6-d$ physical internal dimensions. The coordinate module corresponds to the leftmost node. The line of $n-1$ nodes is the gravity line, giving a solution to the section constraint. $n=0$ corresponds to deleting the ``$GL(1,\RR)$ node(s)" immediately connected to the gravity line, which reveals the $GL(n,\RR)\times Spin(1,\nu+1)$ subgroups, accompanied by the $SU(2)$ ``R-symmetry''  for $\nu=4$. The $U(1)$ for $\nu=2$ is the compact Cartan element of the leftmost pair connected by arrows.
\label{SatakeFigure}}  
\end{narrow}

\end{center}

\end{figure}

Only certain arrangements of black/white nodes and arrows are admitted in a Satake diagram. We will not give a full list, nor try to argue for it. 
It follows from the rules that extending a Satake diagram $\Delta(\fg)$ by attaching white nodes to white nodes leads to a Satake diagram for an extended real Lie algebra with $\fg$ as a subalgebra.  
The diagrams relevant for the magical supergravities are listed in Figure \ref{SatakeFigure}.

\tikzset{/Dynkin diagram/fold style/.style={stealth-stealth,shorten <=1.5mm,shorten >=1.5mm}}

\subsection{Tensor hierarchy algebras and real forms\label{THAsection}}

Tensor hierarchy algebras \cite{Palmkvist:2013vya} are Lie superalgebras, typically infinite-dimensional, that encode the field (and ghost) content of extended geometry (see Section \ref{sec:ExtGeom}).

Given a Lie algebra $\fg_\CC$ (possibly an infinite-dimensional Kac--Moody algebra, but for our purposes a finite-dimensional semi-simple Lie algebra) and a dominant integral weight $\lambda$, 
tensor hierarchy algebras $S(\fg_\CC,\lambda)$ and $W(\fg_\CC,\lambda)$ over $\CC$ are constructed the usual way
\cite{Palmkvist:2013vya,Carbone:2018xqq,Cederwall:2019qnw,Cederwall:2021ymp,Cederwall:2022oyb}.
They are associated with a Dynkin diagram where a ``grey'' (fermionic) node is attached to the Dynkin diagram of $\fg_\CC$,
$\Delta(\fg_\CC)$, according to the decomposition of $\lambda$ in terms of fundamental weights. 
In the examples relevant to us, $\lambda$ is a fundamental weight dual to a simple root at one end of $\Delta(\fg_\CC)$, and we will simply write $S(\fg_\CC)$ and $W(\fg_\CC)$.
Though the tensor hierarchy algebras infinite-dimensional, each degree in a grading 
with respect to the fermionic root is finite-dimensional module of $\fg_\CC$.

Both $S(\fg_\CC)$ and $W(\fg_\CC)$ contain the lowest weight module $R_1=R(-\lambda)$ at degree $1$ and 
$R_2=\vee^2 R(-\lambda)\ominus R(-2\lambda)$ at degree $2$.
In $S(\fg_\CC)$, degree $0$ consists of $\fg$, while degree $-1$ contains all modules that ``automatically'' respect the ideal
$R(-2\lambda)$ at degree $2$, in the sense that $R_1\otimes R_{-1}\supset\fg$ but $R_{-1}\otimes R(-2\lambda)\not\supset R_1$. 
$R_{-1}$ is the embedding tensor module.
In $W(\fg_\CC)$, also a grading element is present at degree $0$ and a module $R(\lambda)$ at degree $-1$.

In refs. \cite{Carbone:2018xqq,Cederwall:2019qnw,Cederwall:2021ymp}, generators and relations analogous to 
the Chevalley--Serre construction were used to define tensor hierarchy algebras.
Taking these generators as generators of a {\it real} superalgebra leads to a real form $S(\fg,\lambda)$ or $W(\fg,\lambda)$ which we call the split real form. At degree $0$, the split real form $\fg$ is found, at level $1$ the real module $R(-\lambda)$, etc.

\begin{figure}[H]
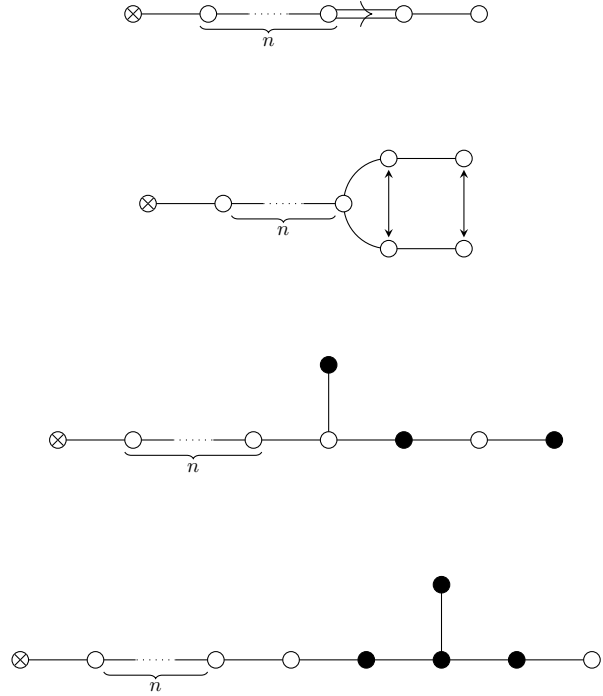

\begin{center}
\begin{dynkinDiagram}[extended,affine mark=t,backwards=true,make indefinite edge={7-8}]E[1]{o****ooo}
\dynkinBrace[n]87
\end{dynkinDiagram}

\begin{narrow}[both=1cm]
\caption{\it Satake diagrams 
of the tensor hierarchy algebras used for the magical supergravities with $n=6-d$ physical internal dimensions. \label{SatakeTHAFigure}}  
\end{narrow}

\end{center}

\end{figure}

In order to define a real form of a tensor hierarchy algebra we need to specify a real form $\fg$ of $\fg_\CC$, with the condition that $R(-\lambda)$ is a real representation. We are then guaranteed that the modules appearing at all degrees
are real $\fg$-modules.
The real tensor hierarchy algebras relevant to the magical supergravities can be described by Satake diagrams
obtained by first extending diagrams of the types in
Figure \ref{SatakeFigure} with a white node $0$ to the left, resulting in a Satake diagram for a real form of $\fg^+$, the next diagram in the series, and then
with a grey node ($\otimes$), numbered $-1$, to the left.
The resulting Satake diagrams associated with real forms $S(\fg^+)$ of the tensor hierarchy algebras are listed
in Figure \ref{SatakeTHAFigure}.
From the diagram one can then define the involution $\sigma$ on the corresponding complex tensor hierarchy algebra,
which in turn defines the real form,
in the same way as for
$\fg_{\mathbb{C}}$.

The involution acts trivially on the generators associated to the white node first added to the Satake diagram of $\fg$
but not on all generators associated to the grey node. This is due to 
a fundamental difference between the tensor hierarchy algebras and the contragredient
Lie superalgebras $\mathscr{B}(\fg^+)$ of Borcherds--Kac--Moody type that are described by the same diagrams, where there is only one generator $f_{-1}$ at degree $-1$.
On the other hand, in $S(\fg^+)$, there is one generator $f_{{-1},i}$ for each node $i$ in the Satake diagram of $\fg$. Under the involution $\sigma$,
these generators transform in the same way as the corresponding Cartan generators $h_i$.
Considering the contragredient
Lie superalgebra, an equivalent diagram is obtained by
extending with \dynkin[edge length=18pt,root-radius=2.5pt]A{tt}
instead of
\dynkin[edge length=18pt,root-radius=2.5pt]A{to}. The corresponding two algebras are isomorphic. By removing the left grey node one then sees that
$\mathscr{B}(\fg)$ is a subalgebra of $\mathscr{B}(\fg^+)$. The corresponding embedding also holds for the tensor hierarchy algebras.

The relevance for the identification of the fields in extended geometry is further detailed in Section \ref{sec:ExtGeom}.
The simplified presentation above holds for tensor hierarchy algebras corresponding to $d\geq3$ (finite-dimensional $\fg$).
Tensor hierarchy algebras corresponding to lower number of external dimensions exhibit more complicated/interesting behaviour, with 
interesting extra modules appearing
\cite{Bossard:2017wxl,Cederwall:2021ymp,Bossard:2021ebg}.

\vspace{4\parskip}
\begin{figure}[H]
\begin{center}
\begin{dynkinDiagram}[extended,affine mark=t,make indefinite edge={1-2}]F{oooo}
\dynkinBrace[n]12
\end{dynkinDiagram}
\end{center}
\end{figure}
%\vspace{12pt}

\begin{figure}[H]
\begin{center}
\begin{dynkinDiagram}[extended,affine mark=t,backwards=true,ply=2,make indefinite edge={2-4}]E{oooooo}
\dynkinBrace[n]24
\end{dynkinDiagram}
\end{center}
\end{figure}

%\vspace{12pt}

\begin{figure}[H]
\begin{center}
\begin{dynkinDiagram}[extended,affine mark=t,make indefinite edge={1-3}]E[1]{o*oo*o*}
\dynkinBrace[n]13
\end{dynkinDiagram}
\end{center}
\end{figure}

%\vspace{12pt}

\section{Extended geometry\label{sec:ExtGeom}}

\subsection{Generalities}

The (real) structure group $G$ with Lie algebra $\fg$ is the continuous version of the duality group.
Let generalised vectors transform in the (real) coordinate representation $R_1=R(-\lambda)$ of $\fg$, which is a lowest weight representation with lowest weight $-\lambda$. This representation is read off from the sequential extensions, \ie, stepwise increment of $n$, of the Satake diagrams of Figure \ref{SatakeFigure}. 
Concretely, the line(s) connecting the leftmost node in the diagram for $g^+$, the algebra obtained by increasing $n$ by $1$, give(s) the Dynkin index for the integral dominant weight $\lambda$.
In tensor notation, we write such a vector $V^M$.

Generalised diffeomorphisms take the usual form \cite{Berman:2012vc} (the ``Dorfman bracket'')
\be
\LL_\xi V^M=\xi^N\*_NV^M+Z_{PQ}{}^{MN}\*_N\xi^PV^Q\;,
\ee
where $Z$ is the invariant tensor \cite{Bossard:2017aae,Cederwall:2017fjm}
\be
Z=\sigma\bigl(-\eta^{\alpha\beta}t_\alpha\otimes t_\beta+(\lambda,\lambda)-1\bigr)\;.\label{eq:Z}
\ee
($\sigma$ is the permutation operator and $\eta$ the inverse Killing metric. Normalisation of roots and weights is chosen such that a long root $\alpha$ has $(\alpha,\alpha)=2$.) 

The commutator of two generalised diffeomorphisms becomes
\begin{align}
[\LL_\xi,\LL_\eta]=\LL_{[\![\xi,\eta]\!]}+\Sigma_{\xi,\eta}\;,
\end{align}
where the ``Courant bracket'' $[\![\cdot,\cdot]\!]$ is the antisymmetrised Dorfman bracket,
\begin{align}
[\![\xi,\eta]\!]={1\over2}(\LL_\xi\eta-\LL_\eta\xi)\;,
\end{align}
and $\Sigma_{\xi,\eta}$ is an ancillary transformation, a section-restricted local $\fg$-transformation. For the purposes of the present letter, it is present only when the number of external dimensions is $d\leq3$. 
This provides the beginning of the $L_\infty$ gauge structure of extended geometry
\cite{Cederwall:2018aab,Cederwall:2019bai}.

The section constraint reads
\be
Y(\*\otimes\*)=0\;,
\ee
where $Y=Z+1$, \ie,
\begin{align}
\sigma Y=-\eta^{\alpha\beta}t_\alpha\otimes t_\beta+(\lambda,\lambda)+\sigma-1\;.\label{eq:generalY}
\end{align}
Concretely, the section constraint expresses the vanishing of all subleading symmetric and antisymmetric modules in the product of 
two derivatives, reflecting the property of the fundamental module of a $GL$ group.

\subsection{Solution of the section constraint\label{SectionSection}}

A section is a linear subspace of the minimal $G$-orbit of $R(\lambda)$ where all vectors $p,q$ satisfy
$Y(p\otimes q)=0$.
%with $\sigma Y=-\eta^{\alpha\beta}t_\alpha\otimes t_\beta+(\lambda,\lambda)-1+\sigma$.
It is well established 
\cite{Bossard:2017aae,Cederwall:2017fjm}
that representatives of such subspaces are obtained by starting from the highest weight state in $R(\lambda)$ (which is a representative in the minimal orbit), and from it sequentially acting with lowering operators associated to negative simple roots along a ``gravity line'' of nodes in the Dynkin diagram. The section then becomes a fundamental $\gl$ module. The explicit form of the $Y$ tensor states the corresponding property of the fundamental $\gl$ module, that the tensor product of it with itself contains a single irreducible module both in the symmetric and antisymmetric parts.
Now it will also be necessary to determine how such solutions behave when the diagram does not consist only of simply laced white nodes. Na\"\i vely,  the gravity line must stop, for example since a compact node does not contribute an $\sl(2,\RR)$ subalgebra. Precise rules are needed for the three cases:
\begin{itemize}
\item{One or two black nodes are encountered;}
\item{A node corresponding to a shorter root is encountered;}
\item{Nodes connected with arrows are encountered.}
\end{itemize}
Given the procedure for solving the section constraint, it is enough to consider subdiagrams of the Satake diagrams containing the different situations. 

When one or two black nodes are encountered, there is always a Satake subdiagram for $\so(1,2m-1)$, with one white node. The section is an isotropic (light-like) subspace of the light c\^one, which is a light ray. The white node is {\it not} part of the gravity line (but its Cartan generator provides the scalings). The gravity line thus ends one step before encountering the black node(s), as in the last two diagrams of Figure \ref{SatakeFigure}.

When a shorter node is encountered, there is an $\sp(4,\RR)\simeq\so(2,3)$ subdiagram. When a pair of nodes connected by arrows is encountered, there is an $\su(2,2)\simeq\so(2,4)$ subdiagram.
In both cases, the maximal isotropic spaces of vectors are $2$-dimensional, so the ``rightmost'' ordinary white node is included in the gravity line, as in the first two diagrams of Figure \ref{SatakeFigure}. The scaling is provided by the node(s) connected to it.

This accounts for the identifications of the gravity lines in Figure \ref{SatakeFigure}.
In all cases, this is of course consistent with the $6$-dimensional origin of the models. It should also be noted that there in all cases is a single $G$-orbit of sections, since no branchings are encountered in the solution of the section constraint.
Similar statements about gravity lines in diagrams for real algebras are found in ref. \cite{Henneaux:2011mm}.

The above statement, that the gravity line runs along any line of simply laced undecorated white nodes, unconnected to black nodes, may also straightforwardly be derived \cite{Sundberg:2022} with the methods of refs. \cite{Bossard:2017aae,Cederwall:2017fjm}. Then one sequentially finds the weights of $R(\lambda)$, starting with the highest one, that spans a solution to the section constraint.
The concrete reason a white node connected to black nodes can not be included in the gravity line is the mixture of the corresponding root with roots of compact $\su(2)$'s under the involution dictating the reality condition.

\tikzset{/Dynkin diagram/fold style/.style={stealth-stealth,shorten <=1.5mm,shorten >=1.5mm}}

\begin{figure}[H]
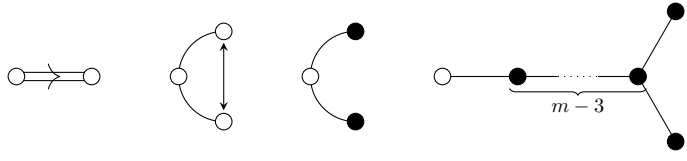


\begin{center}

\begin{dynkinDiagram}B{oo}
\end{dynkinDiagram}
\qquad
%\vspace{1mm}
%\begin{dynkinDiagram}[involutions={13}]A{ooo}
%\end{dynkinDiagram}
\begin{dynkinDiagram}[ply=2,backwards=true]A{ooo}
\end{dynkinDiagram}
\qquad
\begin{dynkinDiagram}[ply=2,backwards=true,arrows=false]A{*o*}
\end{dynkinDiagram}
\qquad
%\vspace{4mm}
\begin{dynkinDiagram}[make indefinite edge={2-3}]D{o****}
\dynkinBrace[m-3]23
\end{dynkinDiagram}

\begin{narrow}[both=1cm]
\caption{{\it Satake subdiagrams for $\sp(4,\RR)$, $\so(2,4)$, $\so(1,5)$  and $\so(1,2m-1)$ ($m\geq4$). The gravity line comes from the left, and includes the leftmost node in the first two cases.}
\label{SatakeSubDiagrams}}  
\end{narrow}

\end{center}

\end{figure}

\subsection{Coset dynamics}

A generalised metric $G_{MN}$ is a symmetric matrix which defines an involution $\tau$ on the Lie algebra through the "transpose" of the representation matrices:
\be
\tau:\,t_{\alpha M}{}^N\mapsto \tilde t_{\alpha M}{}^N=-(G(t_\alpha)^t G^{-1})_M{}^N 
\ee
The involution $\tau$ is in the the same conjugacy class as the Cartan involution $\theta$ of (the real form) $\fg$. 
It has the eigenvalue $1$ on a (locally defined) maximal compact subalgebra $\fk\subset\fg$. 
The "coset 1-form" is 
\be
dGG^{-1}=\Pi^\alpha t_\alpha+\pi \id\;.
\ee
It follows directly %(since $G_{MN}$ is symmetric)
that it has eigenvalue $-1$ under $\tau$, and thus takes values in the orthogonal complement (with respect to the Killing metric) to $\fk$ in $\fg$, $\fk^\perp=\fg\ominus\fk$. 
Note that a scale is included in the metric; we are considering the structure group $G\times\RR^+$. 
The scalings are included in $\fk^\perp$.

When considering only the internal extended geometry, it is convenient to let $G$ have weight $1-2(\lambda,\lambda)$.
Then the internal (pseudo-)Lagrangian density (the ``potential''), invariant under generalised diffeomorphisms, takes the generic form (for $n\leq3$)
\bea
\LL&=G^{MN}\bigl(
\frac12\eta_{\alpha\beta}\Pi^\alpha_M\Pi^\beta_N
-t_{\alpha M}{}^Pt_{\beta N}{}^Q\Pi^\alpha_P\Pi^\beta_Q
+\eta_{\alpha\beta}\ell_{\gamma M}{}^{\beta P}\Pi^\alpha_N\Pi^\gamma_P\bigr.\label{PotentialEq}\\
&\bigl.\qquad-2t_{\alpha M}{}^P\pi_N\Pi^\alpha_P
-{(\lambda,\lambda)\over(\lambda,\lambda)-\frac12}\pi_M\pi_N
\bigl)\;.\nonumber
\eea
The third term contains the invariant tensor $\ell$ appearing among the structure constants for the tensor hierarchy algebra $S(\fg^+)$, where $\fg^+$ is the extension of $\fg$ in the sequences of duality algebras.
It appears only for $d=3$, where $\fg^+$ is an affine algebra. Then, $R(-\lambda)$ is the adjoint representation and 
$\ell_{\alpha\beta}{}^{\gamma\delta}=\eta_{\alpha\beta}\eta^{\gamma\delta}$.

When also external directions are considered, it is convenient to let the determinant $e$ of the external vielbein $e_m{}^a$ assume the r\^ole of the scaling degree of freedom of $G$, so that $dGG^{-1}=\Pi^\alpha t_\alpha$,
The first two terms of eq. \eqref{PotentialEq} can then equivalently, be rewritten as proportional to 
\begin{align}
V_G=e\left(-{k\over4}G^{MN}\*_MG^{KL}\*_NG_{KL}+{1\over2}G^{MN}\*_MG^{KL}\*_LG_{NK}\right)\;,
\end{align}
where $k$ is a constant that will be specified later.

%\subsection{Teleparallel dynamics}
An alternative (equivalent) approach to formulating the dynamics is to use a teleparallel formalism
\cite{Cederwall:2021xqi}. This method is well adapted to the tensor hierarchy algebra, and should be ideal for gauging. 
One uses the torsion $T$ of the 
Weitzenb\"ock connection, taking values in the embedding tensor modules, as a field strength, and the 
Lagrangian contains $T^2$. 

\subsection{Fields from $S(\fg^+)$}

Tensor hierarchies \cite{deWit:2008ta} are an important ingredient in supergravity, as they organise the form gauge fields and their transformations. In ref. \cite{Palmkvist:2013vya}, a class of infinite-dimensional non-contragredient superalgebras, the tensor hierarchy algebras, were constructed, that in a level expansion contain the modules of the form fields, as well as the embedding tensor module. The properties of such algebras were further examined in refs. \cite{Carbone:2018xqq,Cederwall:2019qnw,Cederwall:2021ymp,Cederwall:2022oyb},
and their relation to the symmetries in extended geometry investigated in refs. 
\cite{Palmkvist:2015dea,Cederwall:2018aab,Cederwall:2019bai,Cederwall:2021xqi}.

The content of fields, as well as gauge parameters (ghosts) is thus dictated by the tensor hierarchy algebra
$S(\fg^+)$ \cite{Cederwall:2021ymp}. One introduces a double grading with respect to the two ``leftmost'' nodes, so that the generators at a given bidegree form a module of $\fg$ (which is at bidegree $(0,0)$).
We choose to label the bidegree as $(p,q)$ where $p$ is the level with respect to the second node and $-q$ with respect to the leftmost node in the \dynkin A{tt} extension of Section \ref{THAsection}. The subalgebra $\fg^+$ is found at the line $p=q$.
The degree $\ell$ of the single grading in Section \ref{THAsection} is $\ell=p-q$.

Denote the $\fg$ module found at bidegree $(p,q)$ by $R_{(p,q)}$. Then (for finite-dimensional $\fg$),
$R_{(0,0)}=\adj\oplus{\bf1}$ and $R_{(1,0)}$ is the coordinate module.
The subalgebra at $q=0$ is $W(\fg)\subset S(\fg^+)$ \cite{Cederwall:2019qnw}.
Non-ancillary $p$-form potentials and $(p+1)$-form field strengths are assigned to $R_{(p,0)}$. 
Ancillary $p$-form fields are found in $R_{(p-1,0)}$ precisely when $R_{(p,1)}\neq R_{(p,0)}$. 
See ref. \cite{Cederwall:2019bai} for details.

One of the advantages of the use of an underlying tensor hierarchy algebra is that it reduces the problem of finding fields, gauge transformations etc. to the mathematically more clearly defined problem of constructing a certain superalgebra, also in cases where $\fg$ is infinite-dimensional.

In the following tables, we list the content of a few levels in the tensor hierarchy algebras $S(\fg^+)$ relevant for the magical supergravities with $d=6,5,4,3$. Note the symmetries under $(p,q)\mapsto(d-2-p,1-q)$, signalling the presence of a non-degenerate bilinear form of the superalgebra, and relevant to dualisation in the external dimensions.
Note that more standard orientations of the Dynkin diagrams are used in these tables, rather than the one where $\lambda$ is associated to the leftmost node.

\begin{table}[H]
  \begin{align*}
  \xymatrix@=.4cm{
    \ar@{-}[]+<0.9cm,1em>;[dddd]+<0.9cm,-1em>
    \ar@{-}[]+<-0.8cm,-1em>;[rrrrrr]+<1.4cm,-1em>
&p=-1 &p=0 &p=1&p=2&p=3&p=4\\
q=2&&&&&&{\bf1}\\
q=1&\Theta&\adj&R_1
       &R_2&\bar R_1&\adj\oplus{\bf1}\\
q=0&\bar R_1\oplus\Theta
&{\bf 1}\oplus\adj
           & R_1&R_2&\bar R_1&\adj\\
q=-1 & \bar R_1 & {\bf 1} 
  }
\end{align*}
\vskip-8pt
  \caption{\it Some basis elements of $S(\fg^+)$ for the magical supergravities with $d=6$.}
\label{Sgplus6Table}
\end{table}

\begin{table}[H]
\centering
\begin{tabular}{c|cccc}
&$\nu=1$&$\nu=2$&$\nu=4$&$\nu=8$\\ \hline
\\
$\Delta(\fg)$&\dynkin[edge length=18pt,root-radius=2.5pt]A{o}
	&\dynkin[edge length=18pt,root-radius=2.5pt,involutions=12, edge/.style=white]A{oo}
	&\dynkin[edge length=18pt,root-radius=2.5pt]A{*o*}\quad\dynkin[edge length=18pt,root-radius=2.5pt]A{*}
	&\dynkin[edge length=18pt,root-radius=2.5pt]D{o*****}\\[18pt]
$R_1$&${\bf2}=(1)$&$({\bf2},{\bf1})\oplus({\bf1},{\bf2})$
&$({\bf4},{\bf2})=(100)(1)$&${\bf16}=\DWeight{000}10$\\
&&$=(1)(0)\oplus(0)(1)$\\[8pt]
$R_2$&${\bf3}=(2)$&$({\bf2},{\bf2})=(1)(1)$&$({\bf6},{\bf1})=(010)(0)$&${\bf10}=\DWeight{100}00$\\[8pt]
$\adj$&${\bf3}=(2)$&$({\bf3},{\bf1})\oplus({\bf1},{\bf3})\oplus({\bf1},{\bf1})$
&$({\bf15},{\bf1})\oplus({\bf1},{\bf3})$&${\bf45}=\DWeight{010}00$\\
&&$=(2)(0)\oplus(0)(2)\oplus(0)(0)$
&$=(101)(0)\oplus(000)(2)$\\[8pt]
&&$({\bf3},{\bf2})\oplus({\bf2},{\bf3})$\\
$\Theta$&${\bf4}\oplus{\bf2}$&$ \oplus\,({\bf2},{\bf1})\oplus({\bf2},{\bf1})$
&$({\bf20},{\bf2})\oplus(\bar{\bf4},{\bf2})$
	&${\bf144}=\DWeight{100}10$\\
&$=(3)\oplus(1)$
&$=(2)(1)\oplus(1)(2)$
&$=(110)(1)\oplus(001)(1)$\\
&&$\oplus\,(1)(0)\oplus(0)(1)$
\end{tabular}
        \caption{\it Some representations in the tensor hierarchy algebras for the $d=6$ models.}
    \label{tab:THA6reps}
\end{table}

\begin{table}[H]
  \begin{align*}
  \xymatrix@=.4cm{
    \ar@{-}[]+<0.9cm,1em>;[dddd]+<0.9cm,-1em>
    \ar@{-}[]+<-0.8cm,-1em>;[rrrrr]+<1.4cm,-1em>
&p=-1 &p=0 &p=1&p=2&p=3\\
q=2&&&&&{\bf1}\\
q=1&\Theta&\adj&R_1
       &\bar R_1&\adj\oplus{\bf1}\\
q=0&\bar R_1\oplus\Theta
&{\bf 1}\oplus\adj
           & R_1&\bar R_1&\adj\\
q=-1 & \bar R_1 & {\bf 1} 
  }
\end{align*}
\vskip-8pt
   \caption{\it Some basis elements of $S(\fg^+)$ for the magical supergravities with $d=5$.}
\label{Sgplus5Table}
\end{table}

\begin{table}[H]
\centering
\begin{tabular}{c|cccc}
&$\nu=1$&$\nu=2$&$\nu=4$&$\nu=8$\\ \hline
\\
$\Delta(\fg)$&\dynkin[edge length=18pt,root-radius=2.5pt]A{oo}
%	&\dynkin[edge length=18pt,root-radius=2.5pt,involutions={14;23},indefinite %edge/.style={fill=white,draw=white,color=white},indefinite edge ratio=1]A{oo.oo}
&
%\pgfkeys{/Dynkin diagram,edge length=18pt,root-radius=2.5pt}
\begin{tikzpicture}
\dynkin[name=1,edge length=18pt,root-radius=2.5pt]A{oo}
\node (b) at (0,-.7){};
\dynkin[name=2,at=(b)]A{oo}
\begin{pgfonlayer}{Dynkin behind}
	\foreach \i in {1,...,2}%
	{%
	\draw[/Dynkin diagram/fold style]
	($(1 root \i)$)
 	--
	($(2 root \i)$);%
	}%
\end{pgfonlayer}	
\end{tikzpicture}
	&\dynkin[edge length=18pt,root-radius=2.5pt]A{*o*o*}
	&\dynkin[edge length=18pt,root-radius=2.5pt,backwards=true]E{o****o}\\[18pt]
$R_1$&${\bf6}=(20)$&$({\bf3},{\bf3})=\AAWeight{10}{10}$&${\bf15}=(01000)$&${\bf27}=\EWeight{10}0{00}0$\\[8pt]
$\adj$&${\bf8}=(11)$&$({\bf8},{\bf1})\oplus({\bf1},{\bf8})$
&${\bf35}=(10001)$&${\bf78}=\EWeight{00}0{00}1$\\
&&$=\AAWeight{11}{00}\oplus\AAWeight{00}{11}$\\[8pt]
$\Theta$&${\bf15}\oplus{\bf3}$&$(\bar{\bf3},{\bf6})\oplus({\bf6},\bar{\bf3})\oplus(\bar{\bf3},\bar{\bf3})$
&${\bf105}\oplus\overline{\bf21}$
	&${\bf351'}=\EWeight{01}0{00}0$\\
&$=(21)\oplus(10)$&$=\AAWeight{01}{20}\oplus\AAWeight{20}{01}\oplus\AAWeight{01}{01}$
&=$(10100)\oplus(00002)$
\end{tabular}
        \caption{\it Some representations in the tensor hierarchy algebras for the $d=5$ models.}
    \label{tab:THA5reps}
\end{table}

\begin{table}[H]
  \begin{align*}
  \xymatrix@=.4cm{
    \ar@{-}[]+<0.9cm,1em>;[dddd]+<0.9cm,-1em>
    \ar@{-}[]+<-0.8cm,-1em>;[rrrr]+<1.4cm,-1em>
&p=-1 &p=0 &p=1&p=2\\
q=2&&&&{\bf1}\\
q=1&\Theta&\adj&R_1
       &\adj\oplus{\bf1}\\
q=0&R_1\oplus\Theta
&{\bf 1}\oplus\adj
           & R_1&\adj\\
q=-1 & R_1 & {\bf 1} 
  }
\end{align*}
\vskip-8pt
   \caption{\it Some basis elements of $S(\fg^+)$ for the magical supergravities with $d=4$.}
\label{Sgplus4Table}
\end{table}

\begin{table}[H]
\centering
\begin{tabular}{c|cccc}
&$\nu=1$&$\nu=2$&$\nu=4$&$\nu=8$\\ \hline
\\
$\Delta(\fg)$&\dynkin[edge length=18pt,root-radius=2.5pt]C{ooo}
	&\dynkin[edge length=18pt,root-radius=2.5pt,involutions={15;24}]A{ooooo}
	&\dynkin[edge length=18pt,root-radius=2.5pt]D{*o*oo*}
	&\dynkin[edge length=18pt,root-radius=2.5pt,backwards=true]E{o****oo}\\[18pt]
$R_1$&${\bf14}=(001)$&${\bf20}=(00100)$&${\bf32}=\DWeight{0000}10$&${\bf56}=\EWeight{100}0{00}0$\\[8pt]
$\adj$&${\bf21}=(200)$&${\bf35}=(10001)$&${\bf66}=\DWeight{0100}00$&${\bf133}=\EWeight{000}0{01}0$\\[8pt]
$\Theta$&${\bf64}=(110)$&${\bf70}\oplus\overline{\bf70}$&${\bf352}=\DWeight{1000}01$
	&${\bf912}=\EWeight{000}0{00}1$\\
&&$=(11000)\oplus(00011)$
\end{tabular}
        \caption{\it Some representations in the tensor hierarchy algebras for the $d=4$ models.}
    \label{tab:THA4reps}
\end{table}

%\newpage

\begin{table}[H]
  \begin{align*}
  \xymatrix@=.4cm{
    \ar@{-}[]+<0.9cm,1em>;[dddd]+<0.9cm,-1em>
    \ar@{-}[]+<-0.8cm,-1em>;[rrrr]+<1.4cm,-1em>
&p=-1 &p=0 &p=1&p=2\\
q=2&&&{\bf1}&\adj\\
q=1&\Theta&\adj&\adj\oplus{\bf1}
       &\Theta\oplus\adj\\
q=0&\adj\oplus\Theta
&{\bf 1}\oplus\adj
           & \adj&\Theta\\
q=-1 & \adj & {\bf 1} 
  }
\end{align*}
\vskip-8pt
   \caption{\it Some basis elements of $S(\fg^+)$ for the magical supergravities with $d=3$.}
\label{Sgplus3Table}
\end{table}

\begin{table}[H]
\centering
\begin{tabular}{c|cccc}
&$\nu=1$&$\nu=2$&$\nu=4$&$\nu=8$\\ \hline
\\
$\Delta(\fg)$&\dynkin[edge length=18pt,root-radius=2.5pt]F{oooo}
	&\dynkin[edge length=18pt,root-radius=2.5pt,involutions={16;35}]E{oooooo}
	&\dynkin[edge length=18pt,root-radius=2.5pt,backwards=true]E{o*oo*o*}
	&\dynkin[edge length=18pt,root-radius=2.5pt,backwards=true]E{o****ooo}\\[18pt]
%$R_1$&${\bf14}=(001)$&${\bf20}=(00100)$&${\bf32}=\DWeight{0000}10$&${\bf56}=\EWeight{100}0{00}0$\\[8pt]
$\adj$&${\bf52}=(1000)$&${\bf78}=\EWeight{00}0{00}1$&${\bf133}=\EWeight{000}0{01}0$&${\bf248}=\EWeight{1000}0{00}0$\\[8pt]
$\Theta'$&${\bf324}=(0002)$&${\bf650}=\EWeight{10}0{01}0$
&${\bf1539}=\EWeight{010}0{00}0$
	&${\bf3875}=\EWeight{0000}0{001}0$\\
%&$=(0000)\oplus(0002)$&$=\EWeight{00}0{00}0\oplus\EWeight{10}0{01}0$&$=\EWeight{000}0{00}0\oplus\EWeight{010}0{00}0$
%&$=\EWeight{0000}0{00}0\oplus\EWeight{0000}0{001}0$
\end{tabular}
        \caption{\it Some representations in the tensor hierarchy algebras for the $d=3$ models. The embedding tensor module is $\Theta=\Theta'\oplus{\bf1}$.}
    \label{tab:THA3reps}
\end{table}

\subsection{Extended geometry for magical supergravities}

Note that the Satake  diagrams for the tensor hierarchy algebras in the $\OO$ series, last diagram in Figure
\ref{SatakeTHAFigure}, is a decorated version of the diagram for $S(\fe_{n+6})$, relevant for the extended geometry description of
$D=11$ supergravity with $n+5$ physical internal dimensions.
The two real tensor hierarchy algebras are thus different real forms of the same complex one.

This implies that the dynamics, formulated in terms of a pseudo-action (``pseudo-'' referring to the fact that the section constraint has to be imposed manually, as well as to the self-duality relations occurring for even $d$), takes the same formal expression in the two cases. The extended geometry formulation for $D=11$ supergravity with $d$ external dimensions is well known
for $d=6$ \cite{Abzalov:2015ega}, $d=5$ \cite{Hohm:2013vpa}, $d=4$ \cite{Hohm:2013uia}, $d=3$ \cite{Hohm:2014fxa} 
and $d=2$ \cite{Bossard:2017aae,Bossard:2018utw,Bossard:2021jix}, and of course for $d>6$ \cite{Musaev:2015ces,Hohm:2015xna,Berman:2015rcc}.
Partial results exist for $d=0$ \cite{Bossard:2021ebg}.
However, even if the actions formally look the same, they describe quite different systems, due to the difference in the solutions to the section constraint, which for all versions of the magical supergravities give total physical dimension $6$ (the sum of the number of external dimensions and the dimension of a section).

The extended geometries for the lower $\KK_\nu$ series, $\nu=1,2,4$, are constructed analogously to the ones in the
$\OO$ series. Let us call the algebras appearing in $d=6-n$ algebras of type $\fe_{5+n}$. They have similar sets of invariant tensors, originating in the fact that they are constructed from Jordan algebras over $\KK_\nu$. 
Constructing the analogous actions for the lower series, one needs to identifiy these invariant tensors and the relations they obey, including proper normalisation. 
We will give a concrete example for $d=4$ and algebras of type $\fe_7$.

The only numerical constant that enters the generalised diffeomorphisms is $(\lambda,\lambda)$, the length${}^2$ of the lowest weight in the coordinate representation (the representation of a generalised vector).
It turns out to take the same value for all algebras of the same type, it is thus independent of $\nu$, 
$(\lambda,\lambda)={d-1\over d-2}$. 

\subsection{Example: $d=4$, type $\fe_7$\label{TypeESevenSec}}

As a set of examples, consider the magical supergravities with $d=4$, \ie, with
structure algebras
\begin{align}
\fg=\{\sp(6,\RR),\su(3,3),\so^*(12),\fe_{7(-25)}\}
\end{align}
 for $\nu=1$, $2$, $4$ and $8$ respectively.
The relevant tensor hierarchy algebras are
$S(\fg^+)$, where
\begin{align}
\fg^+=\{\ff_{4(4)},\fe_{6(2)},\fe_{7(-5)},\fe_{8(-24)}\}\;.
\end{align}
These algebras all display the same behaviour, indeed the one expected for models with $d=4$.
In all cases, $\vee^2 R(-\lambda)=R(-2\lambda)\oplus\adj$.
A few degrees are listed in Table \ref{Sgplus4Table}.

The $\fg$-modules appearing in Table \ref{Sgplus4Table} are listed in Table \ref{tab:THA4reps}.
$R_1$ is the coordinate module, which is self-conjugate in these cases.
$\Theta$ is the embedding tensor module.
The presence of a singlet in $R_{(2,1)}$ signals the presence of an ancillary $2$-form.

The $\nu=8$ case for split structure group $E_{7(7)}$ is formulated in ref. \cite{Hohm:2013uia}. 
For the magical models in $d=4$, the structure groups are the conformal groups of the Jordan algebras $J_3(\KK_\nu)$ of hermitean $3\times3$ matrices with elements in $\KK_\nu$. They are $Sp(6,\RR)$, $SU(3,3)$, $SO^*(12)$ and $E_{7(-25)}$, with coordinate modules
%$(001)={\bf14}$, $(00100)={\bf20}$, $\DWeight{0000}01={\bf32}$ and $\EWeight{100}0{00}0={\bf56}$ for $v=1$, $2$, $4$ and $8$ %respectively.
$R_1$ as in Table \ref{tab:THA4reps}.
Thus $\dim R_1=6\nu+8$.
In all cases, $(\lambda,\lambda)={3\over2}$.
The coordinate module is self-conjugate and symplectic; there is an invariant tensor $\Omega_{MN}$, which is used raise fundamental indices by left multiplication. We use the convention $\Omega^{MP}\Omega_{NP}=\delta^M_N$. There is also an invariant symmetric $4$-index tensor, which can be chosen as
$c^{MNPQ}={\mathbb P}^{(MNPQ)}$, where ${\mathbb P}$ is the projector on the adjoint.
The second Casimir operator in the representation $R_1$, $C_2(R(\lambda))={1\over2}(\lambda,\lambda+2\varrho)$, takes the value
$C_2(R_1)={3\over4}(2\nu+3)$.
The projector on the adjoint in $\otimes^2 R_1$ is\footnote{There is a difference in normalisation of the Killing metric compared to ref. \cite{Hohm:2013uia}. We use canonical conventions where the quadratic Casimir operator is
$\hat C_2={1\over2}\eta^{\alpha\beta}t_\alpha t_\beta$, $t_\alpha$ being representation matrices, with $\eta$ normalised so that
it in the adjoint representation becomes ${1\over2}\eta^{\gamma\delta}f_{\gamma\alpha}{}^\epsilon f_{\delta\epsilon}{}^\beta
=g^\vee\delta_\alpha{}^\beta$, \ie, $C_2(\adj)=g^\vee$, the dual Coxeter number.}
\begin{align}
{\mathbb P}_M{}^N{}_K{}^L&=k\eta^{\alpha\beta}t_{\alpha M}{}^Nt_{\beta K}{}^L \label{PttEq}\\
&=k(\eta^{\alpha\beta}t_{\alpha MK}t_\beta{}^{NL}+{1\over2}\delta_M^N\delta_K^L+\delta_M^L\delta_K^N
-{1\over2}\Omega_{MK}\Omega^{NL})\;,\nn
\end{align}
where the constant $k$ takes the values\footnote{The first equality in eq. \eqref{PttEq} holds also in other dimensions, as long as the structure algebra is simple, otherwise more than one constant is needed to form a projection. For $d=5$, $k={2\over\nu+4}$, and for $d=3$, $k={1\over2g^\vee}={1\over6(\nu+2)}$.}
\begin{align}
k={\dim\fg\over2C_2(R_1)\dim R_1}=\{{1\over5},{1\over6},{1\over8},{1\over12}\}={1\over\nu+4}\;.
\end{align}
The section constraint contains the adjoint in the symmetric part of the tensor product and the singlet in the antisymmetric part,
\begin{align}
Y_{MN}{}^{PQ}&=-{1\over k}{\mathbb P}_{MN}{}^{PQ}-{1\over2}\Omega_{MN}\Omega^{PQ}\\
&=-t_{\alpha M}{}^Qt^\alpha{}_N{}^P+{1\over2}\delta_M{}^Q\delta_N^P+\delta_M^P\delta_N^Q
\;.\nn
\end{align}
Notice the relation to eq. \eqref{eq:generalY} with $(\lambda,\lambda)-1={1\over2}$.

The fields needed, in addition to the coset element, are read from the content of the tensor hierarchy algebra
$S(\fg^+)$, Tables \ref{Sgplus4Table} and \ref{tab:THA4reps}. They are: a gauge connections $A_m{}^M$, $2$-forms 
$B_{mn}{}^\alpha$, and also ancillary $2$-forms $B_{mn}{}^M$.
The calculation copies the one in ref. \cite{Hohm:2013uia}, one only needs to keep track of the constant $k$ appearing in various places. We therefore only summarise the results briefly.

The covariant $2$-form field strength is, according to standard tensor hierarchy construction,
\begin{align}
{\mathscr F}_{mn}{}^M=F_{mn}{}^M-{1\over k}t_\alpha{}^{MN}\*_NB_{mn}{}^\alpha-{1\over2}B_{mn}{}^M\;,
\end{align}
where $F$ is constructed through the Courant bracket as
\begin{align}
F_{mn}{}^M=2\*_{[m}A_{n]}{}^M-[\![A_m,A_n]\!]^M\;.
\end{align}
The field strengths are demanded to be selfdual according to
\begin{align}
{\mathscr F}^M+\Omega^{MN}G_{NP}{\star}{\mathscr F}^P=0\;.
\end{align}
There are also field strengths ${\mathscr H}_{mnp}{}^\alpha$ and ${\mathscr H}_{mnp}{}^M$ for the $2$-form fields. They appear in the Bianchi identity for the $2$-form field strength,
\begin{align}
3{\mathscr D}_{[m}{\mathscr F}_{np]}{}^M=-{1\over k}t_\alpha{}^{MN}\*_N{\mathscr H}_{mnp}{}^\alpha
-{1\over2}{\mathscr H}_{mnp}{}^M\;.
\end{align}

The improved Riemann tensor---the improvement needed for Lorentz invariance in the external directions---is
\begin{align}
{\mathscr R}_{mn}{}^{ab}=R_{mn}{}^{ab}(\omega)+{\mathscr F}_{mn}{}^M(e^{-1}\*_Me)^{ab}\;,
\end{align}
where the spin connection is obtained from the vierbein using the covariant derivative
\begin{align}
{\mathscr D}_me_n{}^a=\*_me_n{}^a-A_m{}^M\*_Me_n{}^a-{1\over2}\*_MA_m{}^Me_n{}^a\;.
\end{align}

The full pseudo-Lagrangian density 
\begin{align}
L=L_{\hbox{\tiny EH}}+L_{\hbox{\tiny sc}}+L_{\hbox{\tiny YM}}-V+L_{\hbox{\tiny top}}
\end{align}
then consists of a covariantised Einstein--Hilbert term $L_{\hbox{\tiny EH}}$, a kinetic term for the coset $L_{\hbox{\tiny sc}}$, a Yang--Mills kinetic term $L_{\hbox{\tiny YM}}$,
 a potential term $V$ and a topological term $L_{\hbox{\tiny top}}$. The non-topological terms are
 \begin{align}
 e^{-1}L_{\hbox{\tiny EH}}&={\mathscr R}=e_a{}^me_b{}^n{\mathscr R}_{mn}{}^{ab}\;,\nn\\
 e^{-1}L_{\hbox{\tiny sc}}&={k\over4}g^{mn}{\mathscr D}_mG_{MN}{\mathscr D}_nG^{MN}\;,\nn\\
 e^{-1}L_{\hbox{\tiny YM}}&=-{1\over8}G_{MN}{\mathscr F}_{mn}{}^M{\mathscr F}^{mnN}\;,\\
 e^{-1}V&=-{k\over4}G^{MN}\*_MG^{KL}\*_NG_{KL}+{1\over2}G^{MN}\*_MG^{KL}\*_LG_{NK}\nn\\
 &-{1\over2}g^{-1}\*_Mg\*_NG^{MN}-{1\over4}G^{MN}g^{-1}\*_Mgg^{-1}\*_Ng
 -{1\over4}G^{MN}\*_Mg^{mn}\*_Ng_{mn}
 \nn
 \end{align}
 (the last line in $V$ replaces the terms in eq. \eqref{PotentialEq} containing the scale connection $\pi_M$).
 The topological terms is most conveniently written in terms of 
 integration over a $5$-dimensional manifold with the external $4$-manifold as boundary, so that
\begin{align}
S_{\hbox{\tiny top}}=-{k\over2}\int _{\Sigma_5}d^5x\int [dY]%d^{6\nu+8}Y
\epsilon^{mnpqr}{\mathscr F}_{mn}{}^M{\mathscr D}_p{\mathscr F}_{qrM}
=\int_{\*\Sigma_5}d^4x\int [dY]%d^{6\nu+8}Y
L_{\hbox{\tiny top}}\;.
\end{align}
The internal integration ``$\int[dY]$'' should be seen as purely formal. 
It is not an integral over the $(6\nu+8)$-dimensional internal space, rather over a solution to the section constraint.
This is a pseudo-action with the purpose as a book-keeping device for the equations of motion.
 
 All essential calculations needed to show full invariance of the pseudo-action under internal generalised diffeomorphisms as well as external diffeomorphisms (depending both on external and internal coordinates) have been performed in ref. \cite{Hohm:2013uia}. They require, as usual, cancellations between all terms, and fix the pseudo-action completely.
 The same holds for other values of $d$. For $d=5$, for example, the construction mimics the one in
 ref. \cite{Hohm:2013vpa}. 
In addition to the constant $k$ of eq. \eqref{PttEq}, one will also need to keep track of the normalisation
of the invariant symmetric $3$-index tensor $d^{MNP}$. All fields and algebraic structures are otherwise identical.
%\begin{dynkinDiagram}[make indefinite edge={0-1}]F[1]{oooo}
%\end{dynkinDiagram}

%Also, maybe something more about the tensor hierarchy algebras... gaugings... \&c...

%\section{Examples}

%\subsection{$6$ dimensions: type $E_5$}

%\subsection{$5+1$ dimensions: type $E_6$}

%\subsection{$4+2$ dimensions: type $E_7$}

%\subsection{$3+3$ dimensions: type $E_8$}

\section{Outlook\label{sec:outlook}}

We have demonstrated how extended geometry is formulated for structure groups of arbitrary real forms, with real coordinate modules. The underlying real tensor hierarchy algebra is defined by these data (together with some normalisation when $\lambda$ is not a fundamental weight dual to a long root \cite{Cederwall:2022oyb}). The procedure for solving the section contraint has been explained, resulting in a diagrammatic rule.

It will be straightforward to apply a generalised Scherk--Schwarz reduction 
\cite{Scherk:1979zr,Aldazabal:2013mya,Inverso:2017lrz}
to obtain gauged magical supergravities \cite{Gunaydin:2010fi}.

Coupling to hypermultiplet scalars in the framework of extended geometry presents no further problem, since they are singlets under the structure group. They contribute terms 
\begin{align}
-{1\over2}\sqrt{-g}\left(
g^{mn}G_{AB}D_m\Phi^AD_n\Phi^B+G^{MN}G_{AB}\*_M\Phi^A\*_N\Phi^B
\right)
\end{align}
to the Lagrangian density.
This is relevant for cancellation of anomalies. In particular, it is noteworthy that only  in the (ungauged) $\nu=8$ model coupled to 28 hypermultiplets that
the gravitational anomalies vanish identically.
It may be interesting to understand how an anomalous $6$-dimensional theory is encoded in an extended 
field theory in which the external dimensions are say 3 or 5, where the anomalies must arise from some interplay between external and internal directions.

Our construction only involves the bosonic degrees of freedom. A full supersymmetric version is of course desirable. 
It could use a component field version, with explicit check of the local supersymmetry transformations
as in ref. \cite{Bossard:2019ksx}, or a superfield formulation as in ref. \cite{Butter:2018bkl}. A true extended supergeometry will demand an extension of the structure group itself to a supergroup \cite{Cederwall:2016ukd}.

\begin{figure}[H]
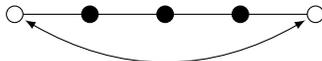

\begin{center}
\begin{dynkinDiagram}[involutions={15}]A{o***o}
\end{dynkinDiagram}
\begin{narrow}[both=1cm]
\caption{\it Satake diagram for $\frak{su}(1,5)$.
\label{SU15Satake}}  
\end{narrow}
\end{center}
\end{figure}

The method can be used to obtain extended geometry formulations of other models, with other homogeneous spaces as scalar cosets.
Just to pick one example without working out the details, let us choose the structure group as $G=SU(1,5)$, Figure \ref{SU15Satake}, and the coordinate module as a $3$-form.
This leades to an extended geometry similar to the $d=4$, $\nu=2$ magical supergravity,
but with a $0$-dimensional section, \ie, with $SU(1,5)$ as R-symmetry. 
Since the coordinate module of $\fg^+$ is 
the adjoint of $E_{6(-14)}$ it should correspond to a $4$-dimensional theory.
Then, the presence of $20$ self-dual gauge fields tell us that this is $D=4$, $N=5$ supergravity \cite{Freedman:2017zgq}.
In a $3+1$ split, the structure group becomes $E_{6(-14)}$, Figure \ref{E6Satake}.

\vskip3\parskip

\newpage
\noindent
\underline{\it Acknowledgments:} This work was initiated during the Mitchell workshop on Exceptional Field Theories, Strings and Holography at Texas A\&M University in April 2018 and finalised during the workshop on Higher Structures, Gravity and Fields at the Mainz Institute for Theoretical Physics
of the DFG Cluster of Excellence PRISMA${}^+$ (Project ID 39083149).
We would like to thank the institute for its hospitality. We would also like to thank Olaf Hohm for taking part in the initial discussions. The work of ES is supported in part by the NSF grant PHYS-2112859.

%\begin{figure}[H]
%\begin{center}
%\begin{dynkinDiagram}[involutions={16}]E{oo***o}
%\end{dynkinDiagram}
%\begin{narrow}[both=1cm]
%\caption{\it Satake diagram for $E_{6(-14)}$.
%\label{E614Satake}}  
%\end{narrow}
%\end{center}
%\end{figure}

%\newpage

\addcontentsline{toc}{section}{References}

\bibliographystyle{utphysmod2}

%\bibliography{biblio}

\providecommand{\href}[2]{#2}\begingroup\raggedright\endgroup

\end{document}